\documentclass[aps,twocolumn,prc,showpacs]{revtex4}
\usepackage{epsfig}
\begin{document}
\title{A doorway to Borromean halo nuclei: the Samba configuration}
\author{M. T.  Yamashita}
\affiliation{Universidade Estadual Paulista, CEP 18409-010 Itapeva, SP, Brasil}
\author{T. Frederico}
\affiliation{Departamento de F\'\i sica, Instituto Tecnol\'ogico de Aeron\'autica,
Centro T\'ecnico Aeroespacial, 12228-900 S\~ao Jos\'e dos Campos, Brasil}
\author{M. S. Hussein}
\affiliation{Instituto de F\'{\i}sica, Universidade de S\~{a}o Paulo,
C.P. 66318, CEP 05315-970 S\~{a}o Paulo, Brasil}
\date{\today}
\begin{abstract}
We exploit the possibility of new configurations in three-body
halo nuclei - Samba type - (the neutron-core form a bound system)
as a doorway to Borromean systems. The nuclei $^{12}$Be, $^{15}$B,
$^{23}$N and $^{27}$F are of such nature, in particular $^{23}$N
with a half-life of 37.7 s and a halo radius of 6.07 fm is an
excellent example of Samba-halo configuration. The fusion below
the barrier of the Samba halo nuclei with heavy targets could
reveal the so far elusive enhancement and a dominance of
one-neutron over two-neutron transfers, in contrast to what was
found recently for the Borromean halo nucleus $^6$He + $^{238}$U.
\end{abstract}
\pacs{25.70.Jj, 25.70.Mn, 24.10.Eq, 21.60.-n}

\maketitle

Borromean nuclei, be them halo or not, are quite common and their
study has been intensive \cite{JeRMP04,BeNSc01,HaARNPS95,VaPS00}.
These three-body systems have the property that any one of their
two-body subsystems is unbound. The halo-type Borromean nuclei are
of special interest as they are unstable and their radii are quite
large compared to neighboring stable nuclei. Typical cases are
$^{11}$Li \cite{Tani} and $^6$He \cite{NaPRL02}. The cores, $c$,
$^4$He and $^9$Li are bound but the three two-body subsystems are
not; $^5$He and $^{10}$Li and the $nn$ system ($n$ represents a
neutron). Borromean excited states of stable nuclei also exist. A
well known example is the Hoyle resonance in $^{12}$C at an
excitation energy of about 6.8 MeV. This resonance, of paramount
importance in stellar nucleo-synthesis, is a cluster of three $^4$He
nuclei, with the unbound $^8$Be as two-body subsystems.

An important issue that has to be addressed is how the halo
Borromean nuclei are formed as more neutrons are added to a given
stable nucleus. In order to answer this question we have made a
survey of the isotopes of several nuclei in the proton $p$-shell
region as well as the fluorine isotopes. We have discovered that the
halo develops in the Borromean nuclei in a gradual fashion. Further,
to reach the halo Borromean nucleus the system, two neutrons down,
acquires a new configuration which in several cases is a halo-type
as well. This new type of halo nuclei has the feature that only one
of its two-body subsystems, the di-neutron, is unbound. The other
two subsystems are weakly bound. Another feature of this
configuration is that the one neutron separation energy, $E_n$, is
smaller than that of the two-neutron, $E_{2n}$, in contrast to the
Borromean halo nuclei where $E_n>E_{2n}$. Of course the situation
$E_n<E_{2n}$ that prevails in the new ``doorway" configuration is
shared by the other normal isotopes. In what follows we shall use
the available information about the isotopes studied here contained
in the Nuclear Wallet Cards (Sixth edition, 2000). Note that no
Borromean isotope exists in oxygen (see, however, Ref.
\cite{JeRMP04}).

As an example we consider the boron isotopes: $A$ = 8, 9, 10, 11,
12, 13, 14, 15, 17, 19. Both $^{17}$B and $^{19}$B would be
Borromean halo nuclei ($^{19}$B more so). The isotope $^{15}$B is
the doorway configuration. The halo radius is 5.15 fm while nuclear
radius is 2.91 fm, to be compared to 2.50 fm of normal $A$ = 15
nucleus. The details of our radius calculation are given below. We
have identified five candidates for  Samba halo nuclei
\cite{raios,tese} of the doorway type. They are $^{12}$Be, $^{15}$B,
$^{20}$C, $^{23}$N and $^{27}$F. The name Samba partly is inspired
by the work of Robicheaux \cite{RoPRA99} who exploited yet another
type of halo three-body systems, such as the hypertriton
$^3_\Lambda$H$_1$, where only one of the two-body subsystems is
tightly bound, and called such systems Tango halo systems. Whereas,
the Tango halo nucleus has a bound two-body subsystem that has to
move in a rather ``harmonic" fashion in the presence of the core,
the Samba nucleus (two of the two-body subsystems are tightly bound)
is distinguished by the fact that the motion of the two-neutron
subsystem can be a rather more ``agitated" and the three-body system
remains bound.

We have calculated the reduced dipole transition strengths $B(E1)$
and the radii of these five candidates for Samba halo nuclei. The
$B(E1)$'s were calculated using a simple cluster model usually
employed to get a rough estimate \cite{BeNPA91}. This model treats
the two neutrons as a cluster that vibrates against the core. A
Yukawa type wave function is used to describe the ground state and a
plane wave is employed for the final continuum state in the
calculation of the dipole matrix element. The model allows for the
derivation of a simple analytical formula for the dipole strength
distribution, $dB(E1)/dE^*$. The dipole distribution is an important
quantity in the study of exotic nuclei that is usually measured
through the electromagnetic dissociation of these fragile systems in
the field of a heavy target such as $^{208}$Pb. Integrating
$dB(E1)/dE^*$ over $E^*$ gives the $B(E1)$ value. The cluster model
gives a simple formula for this
\begin{equation}
B(E1)=\frac{3e^2}{16\pi\mu}\left(\frac{2Z}{A}\right)^2\frac{1}{E_{2n}}, 
\label{be1}
\end{equation}
where $\mu$ is the reduced mass of the core and the $2n$ cluster.
The factor $2Z/A$ corresponds to the number of neutrons in the
halo, 2; the charge of the core, $Z$ and the mass number of the
whole halo nucleus, $A$. Note that the two neutron separation 
energy is inversely proportional to the average distance between 
the core center and the three-body center-of-mass ($CM$), $\sqrt{\langle
r^2_{c-CM}\rangle}$, which is used as the healing distance of the cluster 
model wave function. For, e.g., $^{12}$Be we find for $B(E1)$ the value 
0.043 $e^2$fm$^2$ (another calculation gives 0.05-0.06 $e^2$fm$^2$\cite{nunes}) to
be compared with 0.051(13)~$e^2$fm$^2$ for the experimental
value~\cite{iwasaki}   and to 0.61~$e^2$fm$^2$ for the well
developed halo in the Borromean nucleus $^6$He. Thus one should be
able to get reasonably reliable and appreciable dissociation cross
section for this Samba nucleus (the yield or production cross
section of these nuclei should be larger than those of the
Borromean halo nuclei \cite{CaPRC92}, rendering the experiment
quite feasible). It is worth mentioning here that for Tango halo
nuclei, where the halo is a pair of proton-neutron, the B(E1) is
similar to Eq. (\ref{be1}) except for the energy $E_{2n}$, which
is replaced by $E_{pn}$ and the factor $(2Z/A)^2$ which is
replaced by $[(A-2Z)/A]^2$. This will render the $B(E1)$ for Tango
halo nuclei a factor $[(A-2Z)/2Z]^2$ smaller than that of a
corresponding Samba or Borromean nucleus if $E_{2n}$ is maintained
equal to $E_{pn}$.

\begin{figure}[htb!]
\centerline{\epsfig{figure=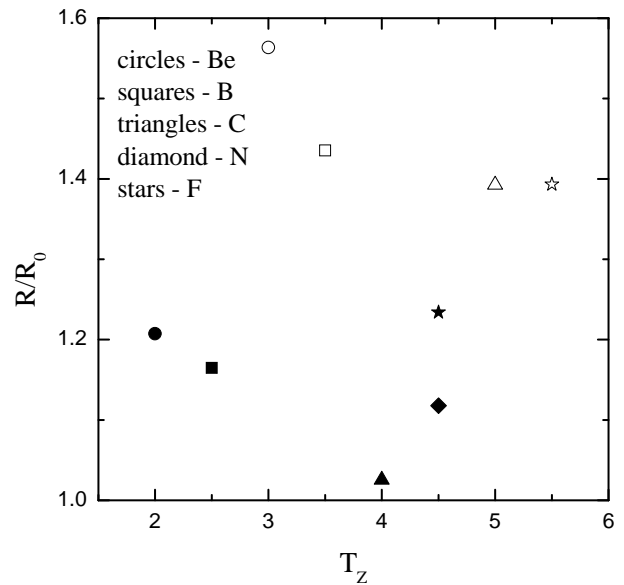,width=8cm}}
\caption[dummy0]{Halo radii of our candidates for two-neutron halo
nuclei in isotopes indicated in the legend. The full symbols are the Samba 
type nuclei (results from Table \ref{radii}) 
and the open symbols are the Borromean nuclei. All the points are divided 
by the the isotope radii, $R_0$, calculated assuming a normal nature of the 
isotope. $T_z$ is the isospin projection.} \label{Be}
\end{figure}

In figure (\ref{Be}) we show the halo radii of our
candidates for two-neutron halo nuclei as a function of the
isospin projection $T_z=(N-Z)/2$. These radii can be estimated
from the neutron-$CM$ root mean square radius $\sqrt{\langle
r^2_{n-CM}\rangle}$ estimated as
$\rho\sqrt{\frac{\hbar^2}{m_nS_3}}$ (see Refs. \cite{raios} and
\cite{tese}). The quantity rho is adimensional. For our purposes $\rho$ is obtained 
calculating the average $(R(^{14}Be)+R(^{17}B))/2$ and associating this to squareroot 
$\sqrt{\langle r^2_{n-CM}\rangle}$ above. Here $R(^{14}Be)=3.74$ fm and $R(^{17}B)=3.8$ fm, 
from Ref. \cite{BeNSc01}, $S_3$ is the three-body energy with respect to the
binding energy of the neutron in the two-body subsystem,
$S_3=E_{2n}(A)-E_n(A-1)$ for Samba nuclei, where $E_n(A-1)$ is the
neutron separation energy in the bound $A-1$ system, and
$S_3=E_{2n}$ for Borromean nuclei. With this value of $\rho$ we calculate all the 
radii shown in the table and figure. The full symbols (see Table
\ref{radii}) are the radii of Samba type halo nuclei calculated
exactly within the three-body model described below and the open
symbols are the radii of the Borromean halo nuclei calculated as
above with $\rho$ set equal 1.35. Both radii appear divided by the
radii calculated assuming a normal nature of the isotopes, namely,
$R_0=1.013A^{1/3}$ fm (the factor 1.013 fm was taken to reproduce
the experimental radius of $^{12}$C of R= 2.32 fm
\cite{tostevin}).

Conspicuous in the figure is lack of indication of one-neutron halo
nuclei whose radii are quite large and deviant from $R_0$. These nuclei, such as $^{11}$Be,
$^{19}$C and others have been shown to have a well developed one-neutron halo. We
did not show these halo nuclei as the thrust of our work here is on three-body,
two-neutron halo nuclei. Also absent in the figure is the, what would be,
Borromean halo nucleus following the Samba one, $^{23}$N. We are tempted to predict
that such a nucleus, $^{25}$N, may exist, though we have no information about $E_n$, $E_{2n}$
and its life time. From the above considerations and we can clearly rule out $^{20}$C as a 
Samba halo nucleus and accordingly $^{22}$C as a Borromean halo nucleus.

The ``doorway" aspect of the Samba halo nuclei is quite evident
especially in the circles, squares and stars. They always
precedes the final Borromean halo nucleus in the chain of
isotopes. It would be indeed very interesting to perform Coulomb
dissociation experiment on, e.g. $^{23}$N, and $^{27}$F to
investigate the dipole strength distribution and also the
longitudinal momentum distribution to assess the halo nature of
these Samba systems. The nucleus $^{12}$Be has a rather subtle
shell model structure with the ground state containing
$(1s_{1/2})^2$, $(1p_{1/2})^2$ and possibly $(1d_{5/2})^2$
configurations, making it less likely to be a clear cut Samba type
halo nucleus.

In Table \ref{radii} we have collected several physical quantities
of  the following Samba type nuclei (treated as a $n-n-c$ three-body
system): $^{12}$Be, $^{15}$B, $^{20}$C, $^{23}$N and $^{27}$F.  The
lower frame shows the results for the core radius
$R_{core}=1.013(A-2)^{1/3}$, the total radius
\begin{equation}
R\sim\sqrt{\frac2A\langle r^2_{n-CM}\rangle+\frac{A-2}{A}R^2_{core}},
\label{rt}
\end{equation}
B(E1,$0\rightarrow1$) (eq. (\ref{be1})), and the half-life, $T_{1/2}$, of the nuclei.

\begin{table}[htb!]
\begin{tabular}{|c|c|c|c|c|} \hline
$nucleus$ & $E_{n}$ & $E_{2n}$ & $\sqrt{\langle r^2_{n-CM}\rangle}$ &
$R_0$ \\
&(MeV)&(MeV)&(fm)& (fm) \\
\hline \hline
$^{12}$Be   & 0.504 & 3.669 & 4.81 & 2.75   \\
$^{15}$B    & 0.973 & 3.734 & 5.15 & 2.96   \\
$^{20}$C    & 0.191 & 3.462 & 4.00 & 3.26   \\
$^{23}$N    & 1.200 & 3.672 & 6.07 & 3.41   \\
$^{27}$F    & 1.041 & 2.412 & 8.94 & 3.60  \\
\hline \hline
$nucleus$ & $R_{core}$ & $R$ &
B(E1,$0\rightarrow1$) & T$_{1/2}$ \\
& (fm) & (fm) & ($e^2$fm$^2$) &\\
\hline\hline
$^{12}$Be   & 2.18 & 2.80 & 0.043 & 21.3 ms  \\
$^{15}$B    & 2.38 & 2.91 & 0.037 & 9.87 ms  \\
$^{20}$C    & 2.66 & 2.82 & 0.013 & 14 ms  \\
$^{23}$N    & 2.80 & 3.22 & 0.024 & 37.7 s  \\
$^{27}$F    & 2.96 & 3.75 & 0.051 & $>$200 ns \\
\hline
\end{tabular}
\caption[dummy0]{Physical quantities of the Samba halo nuclei given
in the first column. The second and third columns are, respectively,
the $n-c$ and the $n-n-c$ energies used to calculate the $n-CM$ root
mean-square radii, $\sqrt{\langle r^2_{n-CM}\rangle}$. $R_0$ is the
nucleus radius and $R_{core}$ is the core radius, $R$ (eq.
(\ref{rt})) is the average between $R_0$ and $\sqrt{\langle
r^2_{n-CM}\rangle}$. B(E1,$0\rightarrow1$) is given by eq.
(\ref{be1}).  $T_{1/2}$ is the nucleus half-life.} \label{radii}
\end{table}

The calculated radii of the Samba nuclei in the three-body model are
a bit larger than the measured one (e.g. for $^{12}$Be, R$_{exp}$ =
2.59 $\pm$ 0.06 fm see Ref. \cite{ozawa}). We trace this small
discrepancy to the neglect of Pauli blocking effect which tends to
make the nuclear potential between the core and the neutrons less
attractive at short distances. In order to not change the three-body
binding energy the system needs to shrink a little to better feel
the nuclear attraction.

The radii in Table I were calculated using the three-body formalism
of Refs. \cite{raios,virtual}. In these references, subtracted
Faddeev equations for the three-body system $n-n-c$ are used.  The
subtraction energy which is required in the model is taken to be
$\mu^2_{(3)}$ ($E=-\mu^2_{(3)}$ is an arbitrary subtraction point
where the $T$-matrix (from this point a small $t$ will be used when 
we refer to the two-body $t$-matrix and a capital one when we refer to 
the three-body $T$-matrix), $T(-\mu^2_{(3)})$, should be known. A more
detailed description about our subtraction method can be found in
Refs. \cite{AdPRL95}). The motivation behind the subtraction is the
use in the model of a delta function potential for the $nn$
interaction which yields divergent two-body $t$-matrix at large
momenta (short distances). Thus the Faddeev equations for the
three-body system must be appropriately subtracted. The $n-c$
interaction is also taken to be of a zero range. Therefore, in this
model the only physical scales used as input are directly related to
observables: the $nn$ scattering length, the energies $E_n$ and
$E_{2n}$. It is worth mentioning that the subtracted three-body
equations are obtained through an elimination procedure that
involves the renormalization of the delta function interaction
strength so that the large momentum divergence alluded to above is
removed. In a nut shell, one takes the two-body matrix $t(E)$ for
the delta interaction, $\lambda\delta(\vec{r}-\vec{r}\,^\prime)$, at
an energy, $-\mu^2_{(2)}$, and identify it with the renormalized
coupling strength $\lambda_R(-\mu^2_{(2)})$. The delta potential
$t$-matrix at any energy can be obtained through usual
Lippmann-Schwinger manipulations as
\begin{equation}
t_R(E)^{-1}=\lambda_R(-\mu^2_{(2)})^{-1}+2\pi^2(\mu_{(2)}+ik).
\end{equation}
where the index $R$ will refer to a renormalized $t$- or $T$- matrix.

The three-body $T$-matrix can be handled in a similar fashion with
the above subtracted two-body matrix. The subtracted three-body
$T$-matrix can be written as
\begin{eqnarray}
\nonumber
T_R(E)=&&T_R(-\mu^2_{(3)})+T_R(-\mu^2_{(3)})\left[G_0(E)\right.\\
&&\left.-G_0(-\mu^2_{(3)})\right]T_R(E).
\label{T}
\end{eqnarray}

It is a simple matter to show that $T_R$ above is independent on the value of the
subtraction energy $-\mu^2_{(3)}$, namely $dT(E)/d\mu^2_{(3)}=0$. This is clear by
construction. Formally, this independence of $T(E)$ on $\mu_{(3)}$ can be shown by
resorting to the identity, $dT(E)/d(E)=-T(E)G^2_0(E)T(E)$, which is clearly also valid for
$dT(-\mu^2_{(3)})/d\mu^2_{(3)}$. Thus we have, using Eq. (\ref{T}),
\begin{equation}
\frac{dT_R(E)}{d\mu^2_{(3)}}=2\mu_{(3)}\frac{dT_R(E)}{d\mu_{(3)}}=0.
\end{equation}
The second relation in the above formula is a Callan-Symanzik type, commonly
used in renormalization procedure of field theories.

Armed with the above invariance of $T_R(E)$ with regards to the
subtraction energy and the removal of the ultraviolet divergence of
the delta potential two-body $t$-matrix, the three-body $T$-matrix
can be used to construct the corresponding wave function needed to
calculate matrix elements among which is the root mean-square radius
of a given nucleus \cite{raios}. The mean-square distances ($\langle
r^2_{n-CM}\rangle$ and $\langle r^2_{c-CM}\rangle$) and the coupled
subtracted integral equations for the Faddeev spectator components
are obtained from expressions given in Refs. \cite{raios,tese}.

Before ending, we should mention what is expected of the Samba
halo nuclei when they are used to induce reactions on heavy
targets. In particular, at very low energies (in the vicinity of
the Coulomb barrier) we expect that the Samba halo nucleus to fuse
with, say, $^{208}$Pb, with a probability which is not so much
affected by the breakup coupling, and may exhibit a halo
enhancement, contrary to what was found in the $^6$He$+^{238}$U
studied in Ref. \cite{RaNa04}. In further contrast to $^6$He, the
Samba nucleus would exhibit a one-neutron transfer process
competing with the two-neutron one. The one-neutron transfer cross
section could be larger than the two-neutron one, depending on the
extension of the corresponding configuration in the Samba halo
nuclei. A specific Samba nucleus that we suggest to investigate
experimentally is $^{23}$N which has a life time of 37.7 s and a
halo radius of 6.07 fm. The enhancement of the fusion, sought for
in vain in Borromean nuclei~\cite{RaNa04,HuPRC92}, may come into
light with the Samba nuclei.

\noindent
Acknowledgements: We thank Professor P.G. Hansen for comments and suggestions.
The work of MTY is suppported by FAPESP, that of TF partly by FAPESP and the
CNPq and that of MSH partly by FAPESP, the CNPq and the Instituto do Mil\^enio de Informa\c{c}\~ao
Qu\^antica-MCT, all Brazilian agencies.

\end{document}